\documentstyle[12pt,aps]{revtex}%

\newcommand{\be}{\begin{equation}}
\newcommand{\ee}{\end{equation}}
\newcommand{\bea}{\vspace{0.25cm}\begin{eqnarray}}
\newcommand{\eea}{\end{eqnarray}}

\def\PLA{{Phys. Lett.}  A }
\def\PLB{{Phys. Lett.}  B }
\def\PRL{{Phys. Rev. Lett.} }
\def\PRA{{Phys. Rev.} A }

\def\PRD{{Phys. Rev.} D }

\begin{document}

\centerline{\bf Review of studies about quantum communication }
\centerline{\bf and
foundations of quantum mechanics at IENGF }
\vskip 1cm
 \centerline{ Marco
Genovese }

\centerline{ Istituto Elettrotecnico Nazionale Galileo Ferraris,
Strada delle Cacce 91, 10135 Torino, Italy \\}


\centerline{\bf Abstract}

In this proceeding I review the main experimental results obtained at IENGF (Turin, Italy) by using a source of entangled photons realised superposing, by means of an optical condenser, type I PDC produced in two crystals. More in details, after having described how this source is built, I will report on a Bell inequalities test obtained with it.
Then I describe a recent innovative double slit experiment realised with a similar scheme.
Finally, I hint about future developments of this activity at IENGF and in particular about a quantum cryptographic scheme in d=4.

\section{Introduction}

\noindent In the last years entanglement has been recognised as a
main resource for quantum communication and quantum computation
\cite{NC}. In quantum communication the use of entangled photons
is, without any doubts, the main resource for future developments
\cite{NC}.

Among the most fascinating experiments on the possibilities
offered by sharing quantum entanglement are quantum teleportation
\cite{tele}, quantum dense coding \cite{dc} and entanglement
swapping \cite{sw}. Moreover, due to the impossibility of
intercepting a quantum state without disturbing the transmission
\cite{NC}, secure quantum key distribution is possible. Quantum
cryptographic transmissions of many tens kilometres have already
been realised \cite{lontano} and this technology is approaching an
applicative stage.

Furthermore, the generation of entangled pairs of photons has been and is of the largest relevance for experiments addressed to test foundations of quantum mechanics \cite{Aul}.

Thus the realisation of efficient sources of entangled states is of the utmost relevance.

Various schemes have been proposed for generating entangled states of photons.

The first schemes were based on the use of polarisation entangled states of photons generated by cascade decay of atoms \cite{asp}, but due to the atom recoil the angular correlation of the produced photons was poor and very low detection efficiencies were reached.

In the 90's a big progress in the direction of larger efficiencies has been obtained by using parametric down conversion (PDC) processes \cite {Mandel}, for
PDC presents angular correlations better than 1 mrad. The first experiments \cite{type1} using this technique, where performed with type I PDC (where the two photons have the same polarisation), which gives phase and momentum entanglement. These schemes represent the better solution for long distance communication in fibers \cite{lontano}.However, on the other hand,
the use of beam splitters strongly reduces the total quantum efficiency.

In alternative, a polarisation entangled state can be generated \cite{ou}. However, in generating this state, in the first
configurations,  half of the initial photon flux was lost.

More recently,  polarisation entangled states of two photons have
been produced by Type II PDC \cite{type2} or by superimposing two
type I PDC emissions, in this case or of two thin adjacent
crystals \cite{whi} either by inserting them in an interferometer
\cite{shih}. These schemes allow the realisation of rather
efficient sources.

Incidentally, also polarisation entangled states of three
\cite{3e} or four \cite{4e}  photons were produced, but with very
low efficiency.

In this proceeding I describe a source where two type I emissions are also superimposed, but by using an optical
element between the two crystals. The advantages of this scheme are that, in principle, a perfect superposition can
be obtained and that non-maximally entangled states of different degrees of entanglement can be easily realised.
I will then discuss some applications of this set-up, or of some simple modifications of it, to studies concerning
quantum communication and foundations of quantum mechanics.

\centerline{Our source}

The general scheme of our set-up
is shown in Fig. 1: two crystals of $LiIO_3$ (10x10x10 mm) \footnote{Which we have measured to have $d_{31} = 3.5
\pm 0.4$ \cite{BGN}.} are placed along the pump laser propagation (an argon laser emitting at 351 nm), 250 mm apart,  a distance smaller than the coherence length of the pumping laser. This guarantees indistinguishability in the creation of a couple of photons in the first or in the second crystal. A couple of planoconvex lenses of 120 mm focal length centred in between focalises the spontaneous emission from the first crystal into the second one maintaining exactly, in principle, the angular spread. A hole of 4 mm diameter is drilled into the centre of the lenses to allow transmission of the pump radiation without absorption and, even more important, without adding stray-light, because of fluorescence and diffusion of the UV radiation.   This configuration, which realises  a so-called "optical condenser", has been chosen among others, using an optical simulation program, as a compromise between minimisation of aberrations (mainly spherical and chromatic) and losses due to the number of optical components. The pumping beam at the exit of the first crystal is displaced from its input direction by birefringence: a small quartz plate (5 x5 x5 mm) in front of the first lens of the condensers compensates this displacement, so that the input conditions are prepared to be the same for the two crystals, apart from alignment errors.
Finally, a half-wavelength plate immediately after the condenser rotates the polarisation of the argon laser beam and excites in the second crystal a spontaneous emission cross-polarised with respect to the first one. The dimensions and positions of both plates are carefully chosen in order not to intercept the relevant PDC emission.

Due to the coherent superposition of the two PDC emissions of different polarization the output of this set-up is the state:
\be
\vert \Psi \rangle ={\vert H \rangle \vert H \rangle + f \vert V
\rangle \vert V \rangle \over \sqrt{1 + \vert f \vert ^2}}
\label{PsiH}
\ee
where $H$ and $V$ are horizontal and vertical polarisation respectively.

A very interesting degree of freedom of this configuration is given by the fact that by tuning the pump intensity between the two crystals one can easily
tune (keeping the maximum available power for the first PDC) the value of $f$, which determines how far from a maximally entangled state ($f=1$) the produced state is. This is a fundamental property, which allows to select the most appropriate state for the experiment.

The main difficulty of this configuration is in the alignment,
which is of fundamental importance for having a high visibility.
This problem has been solved using a technique, that  had already
been applied  in our laboratory \cite{JMO} for metrological
studies, namely the use of  an optical amplifier scheme, where a
solid state laser is injected into the crystals together with the
pumping laser. If the angle of injection is selected
appropriately, a stimulated emission along the correlated
direction appears, permitting to identify quite easily two
correlated directions. Then, stopping the stimulated emission of
the first crystal, and rotating the polarisation of the diode
laser one obtains the stimulated emission in the second crystal
and can check the superposition with the former.

This source is also very bright since we observe about 10 kHz
coincidence rate at 200 mW pump power (result next to the best
obtained with two adjacent thin type I crystals and by far larger
than the ones realised with type II sources \cite{kw}). This
property is very important for many different applications as
remote quantum communication and experiments were a high statistic
is needed or where a strong spectral selection strongly reduces
the signal, as when one wants to have interaction about the
entangled photon and specific atomic levels. In particular,
interaction of entangled photons with atomic levels, which
requires very strong spectral selection (a band width around
$10^{-5} nm$), is very important for many applications as quantum
memories \cite{atom}, remote quantum clocks synchronisation
\cite{or}, realisation of quantum logical gates, etc.
Incidentally, in our laboratory a work is in progress (in
collaboration with Camerino University, LENS and INOA) for
realising a controlled-not gate by interaction of two photons by
mean of Kerr effect in Bose-Einstein condensate \cite{cnot}.

Finally, let us notice that by modifying the polarization of one branch or the phase of the parameter $f$ all the four Bell states can be easily generated.

A test of the state produced by this source can be obtained by measuring Bell inequalities, as discussed in the following paragraph.

\centerline{Applications to a test of Bell inequalities}

In 1964 Bell demonstrated \cite{bell} that one can test, with complete generality, standard quantum mechanics (SQM) against every local hidden variable theory (LHVT) by considering specific inequalities involving correlation measurement on entangled states. More recently these inequalities assumed a relevant role in quantum communication as well, since it was shown that they can be used for checking the presence of an eavesdropper in quantum key distribution protocols using entangled states \cite{NC}.

It must also be noticed that the possible existence of a hidden variable theory, of which quantum mechanics would represent a statistical limit, would not only be of exceptional conceptual interest, but would also have utmost consequences on the very foundations of quantum information.

Many experiments have already been devoted to test Bell inequalities
\cite{Mandel,asp,franson,type1,type2,whi}, leading to a substantial agreement with quantum
mechanics and disfavouring realistic local hidden variable theories.
However, due to the low total detection efficiency (the so-called "detection
loophole") no experiment has yet been able to exclude definitively realistic
local hidden variable theories, for it is necessary a further additional
hypothesis \cite{santos}, stating that the observed sample of particles
pairs is a faithful subsample of the whole. This problem is known as  detection or efficiency loophole
 \footnote{Incidentally, the use of equalities \cite{3e,GHZ} instead of Bell inequalities does not change the situation \cite{garuccio}.}.
Incidentally, it must be noticed that a recent experiment
\cite{Win} based on the use of Be ions has reached very high
efficiencies (around 98 \%), but in this case the two subsystems
(the two ions) are not really separated systems during the
measurement and the test cannot be considered a real
implementation of a detection loophole free test of Bell
inequalities, even if it represents a relevant progress in this
sense. Analogously, the suggestion that a loophole free experiment
could be obtained by using K or B mesons \cite{BK} has been shown
to be wrong \cite{nosK}.

  Considering the extreme relevance
of a conclusive elimination of local hidden variable theories, the research
for new experimental configurations able to overcome the detection loophole
is of the greatest interest.

A very important theoretical step in this direction has been achieved
recognising that for non maximally entangled pairs a total
efficiency larger than 0.67 \cite{eb} (in the limit of no background) is required to
obtain an efficiency-loophole free experiment, whilst for maximally
entangled pairs this limit rises to 0.81.

Our set-up, allowing the generation of a chosen non-maximally entangled state, allows a step in this direction.
In particular we have produced a state with $f \simeq 0.4$.

As a first check of our apparatus, we have measured the interference
fringes, varying the setting of one of the polarisers, leaving the other
fixed. We have found a high visibility, $V=0.98 \pm 0.01$, confirming that a good alignment was reached.

As Bell inequality test we have considered the Clauser-Horne sum,

\begin{equation}
CH=N(\theta _{1},\theta _{2})-N(\theta _{1},\theta _{2}^{\prime })+N(\theta
_{1}^{\prime },\theta _{2})+N(\theta _{1}^{\prime },\theta _{2}^{\prime
})-N(\theta _{1}^{\prime },\infty )-N(\infty ,\theta _{2})  \label{eq:CH}
\end{equation}
which is strictly negative for local realistic theory. In
(\ref{eq:CH}), $N(\theta _{1},\theta _{2})$ is the number of
coincidences between channels 1 and 2 when the two polarisers are
rotated to an angle $\theta _{1}$ and $\theta _{2}$ respectively.
Because of low detection efficiency we have substitute in Eq.
\ref{eq:CH}, as in any experiment performed up to now,  single
counts $N(\theta _{1}^{\prime } )$ and $N(\theta _{2})$ with
coincidence counts $N(\theta _{1}^{\prime },\infty )$ and
$N(\infty ,\theta _{2})$, where $\infty $ denotes the absence of
selection of polarisation for that channel. This is one of the
form in which detection loophole manifests itself.

For quantum mechanics $CH$ can be larger than zero, e.g. for a maximally
entangled state the largest value is obtained for $\theta _{1}=67^{o}.5$ , $%
\theta _{2}=45^{o}$, $\theta _{1}^{\prime }=22^{o}.5$ , $\theta _{2}^{\prime
}=0^{o}$.

For non-maximally entangled states the angles for which CH is maximal are
somehow different and the maximum is reduced to a smaller value. The angles corresponding to
the maximum can be evaluated maximising Eq. \ref{eq:CH} with

\bea
\left. \begin{array}{l}

  N[\theta _{1},\theta _{2}] =  [ \epsilon _1^{||} \epsilon _2^{||} (Sin[\theta _{1}]^{2}\cdot Sin[\theta_{2}]^{2}) +
  \epsilon _1^{\perp} \epsilon _2^{\perp}
(Cos[\theta _{1}]^{2} \cdot Cos[\theta _{2}]^{2} )\\
  (\epsilon _1^{\perp} \epsilon _2^{||} Sin[\theta _{1}]^2\cdot Cos[\theta _{2}]^2 + \epsilon _1^{||} \epsilon _2^{\perp}
Cos[\theta _{1}]^2 \cdot Sin [\theta _{2}]^2 )  \\
 + |f|^{2}\ast (\epsilon _1^{\perp} \epsilon _2^{\perp} (Sin[\theta _{1}]^{2}\cdot Sin[\theta_{2}]^{2}) +  \epsilon _1^{||} \epsilon _2^{||}
(Cos[\theta _{1}]^{2} Cos[\theta _{2}]^{2} ) +\\
(\epsilon _1^{||} \epsilon _2^{\perp} Sin[\theta _{1}]^2\cdot
Cos[\theta _{2}]^{2} + \epsilon _1^{\perp} \epsilon _2^{||}
Cos[\theta _{1}]^2 \cdot Sin [\theta _{2}]^2 )   \\
  +  (f+f^{\ast }) ((\epsilon _1^{||} \epsilon _2^{||} + \epsilon _1^{\perp} \epsilon _2^{\perp} - \epsilon _1^{||} \epsilon _2^{\perp} -
\epsilon _1^{\perp} \epsilon _2^{||})
\cdot (Sin[\theta _{1}]\cdot Sin[\theta _{2}]\cdot
Cos[\theta _{1}]\cdot Cos[\theta _{2}]) ]  /(1+|f|^{2}) \,
\end{array}\right. \, .
\label{cc}
\eea
where (for the case of non-ideal polariser) $\epsilon _i^{||}$ and $\epsilon _i^{\perp}$
correspond to the transmission when the polariser (on the branch $i$)  axis is aligned or normal to the polarisation axis respectively.

For our choice $f \simeq 0.4$ the values giving the largest violation of Clauser-Horne inequality are $\theta_1 =72^o.24$, $\theta_2=45^o$, $\theta_1 ^{\prime}= 17^o.76$ and  $\theta_2 ^{\prime}= 0^o$.

In order to measure $CH$ we selected (geometrically and by using two interferential filters of 4 nm FWHM) two conjugated directions at 789 and 633 nm. Photo-detectors were two avalanche photodiodes with active quenching (EG\&G SPCM-AQ) with a sensitive area of 0.025 $mm^2$ and dark count below 50 counts/s.
PDC signal was coupled to an optical fiber (carrying the light on  the detectors) by means of a microscope objective with magnification 20, preceded by a polariser (with extinction ratio $10^{-6}$).

The output signals from the detectors were then routed  to a two channel counter, in order to have the number of events on single channel, and to a
Time to Amplitude Converter circuit, followed by a single channel analyser, for
selecting and counting coincidence events.

Our experimental result is $CH = 513 \pm 25$ coincidences per second,
which is more than 20 standard deviations from zero and compatible with the theoretical value predicted by quantum mechanics.

Thus, our result represents a further indication favouring SQM against LHVT. Its main interest is due to the fact that using tunable non-maximally entangled states is a relevant step toward a conclusive experiment eliminating the detection loophole.

Furthermore, it allows also to exclude some specific local
realistic models. In particular, we have considered the model of
Casado et al. \cite{Santos2}. These authors have presented a local
realistic model addressed to be compatible with all the available
experiments performed for testing local realism. This model
represents the completion of series of papers where this scheme
has been developed \cite{Santosv}. The main idea is that the
probability distribution for the hidden variable is given by the
Wigner function, which is positive for photons experiments.
Furthermore a model of photodetection, which departs from quantum
theory,  is built in order to reproduce available experimental
results.

A great merit of this model is that it gives a number of constraints, which do not follow from the quantum theory and are experimentally testable.

In particular, there is a minimal light signal level which may be reliably detected: a difference from quantum theory is predicted at low detection rates, namely when the single detection rate $R_S $ is lower than

\begin{equation}
R_S < { \eta F^2 R_c^2 \over 2 L d^2 \lambda \sqrt{ \tau T} }
\label{rate}
\end{equation}
where $\eta$ is the detection quantum efficiency, F is the focal distance of the lens in front of detectors, $R_c$ is the radius of the active area of the non-linear medium where entangled photons are generated, $\tau$ is the coherence time of incident photons, d is the distance between the non-linear medium and the photo-detectors, $\lambda$ the average wavelength of detected photons. L and T are two free parameters which are less well determined by the theory \cite{Santosv}: L can be interpreted as the active depth of the detector, while T is the time needed for the photon to be absorbed and should be approximately less than 10 ns \cite{Santosv}, being, in a first approximation, the length of the wave packet divided for the velocity of light.

Referring to the parameters of Eq. \ref{rate} we have $\eta = 0.51 \pm 0.02$ (a value which we have directly measured by using PDC detector calibration \cite{JMO}, see Fig. 3), F= 0.9 cm, $R_c = 1 $ mm, d= 0.75 m and $\tau = 4.2 \cdot 10^{-13} s$ (due to spectral selection by an interferential filter). L can be estimated of $3 \cdot 10^{-5}$ m.
This leads to $T > 1 s$, extremely higher than the limit of 10 ns suggested in the model \cite{Santos2}. Thus, we are strictly in the condition where quantum mechanics predictions are expected to be violated and, in particular, a strong reduction of visibility is expected.
Nevertheless, our results show  a strong violation of Clauser-Horne inequality and a high visibility, in  agreement with standard quantum mechanics, and therefore substantially exclude \cite{nosW} the model of Ref. \cite{Santos2}

\centerline{An innovative double slit experiment}

Even if Bell inequalities experiments will lead to a conclusive test of local hidden variable theories, non-local hidden variable models (NLHVT) will still be possible.

The most interesting example of NLHVT is the de Broglie Bohm theory (dBB).
dBB \cite{7} is a deterministic theory where the hidden variable (determining the evolution of a specific system)
is the position of the particle, which follows a perfectly defined trajectory in its motion.
The evolution of the system is given by classical equations of motion, but an additional potential  must be included.
 This "quantum" potential is related to the wave function of the system and thus it is non-local.
  The inclusion of this term, together with an initial distribution of particle positions given by the quantum
  probability density,  successfully allows the reproduction of {\it almost} all the predictions of quantum mechanics.
  Nevertheless, a possible discrepancy between SQM and dBB in specific cases has been recently suggested by  Ref.s \cite{4,5,6}

In particular Ref.s \cite{4,5,6} suggest that differences can appear in a double slit experiment where two identical particles cross each a specific slit at the same time.

Such a configuration can be easily realised with our set-up substituting the second crystal with a double slit. Although the former theoretical prediction is still somehow subject to discussion \cite{disc}, we think that our results \cite{nosdBB}, in agreement with SQM predictions but at variance with dBB ones, represent a relevant contribution to the debate about the foundations of quantum mechanics urging a final clarification about validity of this theoretical proposal.
More in details, in our set-up (see Fig. 4)  a 351 nm
pump laser of 0.5 W power is directed into a lithium iodate
crystal, where correlated pairs of photons are generated by type I
Parametric Down Conversion. By means of an optical condenser and
within two correlated directions corresponding both to 702 nm
emission (the degenerate emission for a 351 nm pump laser at $2^o$
from the pump in our configuration), the produced photons are sent
on a double slit (obtained by a niobium deposition on a thin glass
by a photolithographic process) placed just before the focus of
the lens system. The two slits are separated by  100  $\mu$m  and
have a width of 10 $\mu$m. They lay in a plane orthogonal to the
incident laser beam and are orthogonal to the table plane. The
distance of the double slit from the focus is of $\approx 1.4 mm$
and therefore the position of the single photon of each pair at
the slit is fixed with a $0.25 \mu m$ precision \cite{joe} (much
smaller than the slit width).

Two single photon detectors are placed at 1.21 m and at 1.5 m
from the slits after an interferential filter at 702
nm, whose full width at half height is 4 nm, and a lens of 6 mm diameter and 25.4 mm focal length.

The output signals from the detectors are routed  to a two channel counter, in
order to have the number of events on a single channel, and to a  Time to
Amplitude Converter (TAC) circuit, followed by a single channel analyser, for
selecting and counting the coincidence events.

The predictions of Ref.s \cite{4,5,6} are that no coincidences
should be observed in the same semiplane respect to the double
slit symmetry axis.

In Fig. 5 we report the measured coincidence pattern. The data are obtained by averaging  7 points of 30' acquisition each. One detector is placed at $-5.5$ cm from the symmetry axis, whilst the second is moved sweeping the whole diffraction peak.
The pattern predicted by SQM significantly agrees with the data. A clear coincidence signal is observed also when the two detectors are placed in the same semiplane respect to the double slit symmetry axis. In particular, when the centre of the lens of the first detector is placed -1.7 cm after  the median symmetry axis  of the two slits (the minus means to the left of the symmetry axis looking towards the crystal) and the second detector is kept at -5.5 cm, with 35 acquisitions of 30' each we obtained 78 $\pm$ 10 coincidences per 30 minutes after background subtraction, ruling out a null result at nearly eight standard deviations.
Thus, if  the former theoretical prediction will be confirmed, this experiment poses a strong constraint on the validity of de Broglie-Bohm theory, which represents the most successful example of a non-local  hidden variable theory

A further interesting property of this scheme is that it allows a new clear test of the connection between which path knowledge and absence of interference.
Since idler and signal photons have no precise phase relation and each photon crosses a well defined slit, no interference appears at single photon detection level. When the coincidence pattern is considered, path undistinguishability is established since the photodectector 1 (2) can be reached either by the photon which crossed slit A or by the one that went through slit B and vice versa. Thus, even if no second order interference is expected, a fourth order interference  modulates the observed diffraction coincidence pattern.

In Fig. 6 we report the observed coincidence pattern (with 10 acquisitions of one hour for each point) obtained when the first detector scans the diffraction pattern, while the second is positioned at $-1$ cm from the symmetry axis. The iris in front of the first detector is of 2 mm. Even if the data have large uncertainties there is a good indication of the fourth-order interference: the interference pattern predicted by SQM fits the data with a reduced $\chi ^ 2$  of $0.9$. By comparison, a linear fit (absence of interference) gives $\chi ^ 2 =12.6$ (with 5 degrees of freedom) and is therefore rejected with a $5 \%$ confidence level. On the other hand we have checked that, as expected, the single channel signal does not show any variation in the same region: the measured ratio between the mobile and the fixed detector is essentially constant (within uncertainties) in this region.

Finally, let us mention that this experiment, as any other experiment concerning a precise manipulation of single quantum states, represents an interesting development for quantum information.

\centerline{Future developments: a set-up for quantum
communication in d=4}

In a recent series of papers \cite{LA} it has been shown how the use of higher dimensions Hilbert spaces for codification can increase the security of a quantum channel.

In the following I expose a proposal of a simple modification of
the former scheme for producing entangled pairs of photons which
allows the codification in $d=4$. Scheme that could be implemented
in a near future in our laboratory.

The change consists in placing on the pump beam a Mach-Zender
interferometer (whose path length difference is large compared to
the pump pulse length) before the non-linear system where a
polarisation entangled pair is generated (see Fig.7).
The pump photon can thus follow the short or the long path
originating the superposition \cite{Gisen}:
\begin{equation}
\vert \Psi_p \rangle = {\frac{ 1 }{\sqrt{2}}} \left [ \vert s \rangle + e^{i
\phi} \vert l \rangle \right ]
\end{equation}
where $\vert s \rangle $ and $\vert l \rangle $ denote the photon which has
followed the short and the long path respectively and $\phi$ the phase
difference between the two paths.

The final bi-photon state is:

\begin{equation}
\vert \Psi \rangle = {\frac{ 1 }{2}} \left [ \vert s H \rangle \vert s H
\rangle + \vert s V \rangle \vert s V \rangle + e^{i \phi} (\vert l V
\rangle \vert l V \rangle + \vert l H \rangle \vert l H \rangle ) \right ]
\label{psi}
\end{equation}
where $H$ and $V$ denote the horizontal and vertical polarisation
respectively, whilst $\vert s \rangle $ and $\vert l \rangle $ denote a
photon created by a pump photon having travelled via the short or the long
arm of the interferometer \footnote{qutrits, d=3, would be generated if the interferometer is inserted between the
 two crystals}.

For implementing quantum communication, one photon is sent to Alice, the
other to Bob. Both select (see Fig. 8) the photon by its polarisation (by using a polarising beam splitter), choosing different bases and then send it to a
Mach-Zender interferometer, which introduces exactly the same difference of
travel times, within the coherence time of the down converted photons,
through the two arms as the interferometer on the pump (see \cite{Gisen}).
Here they can choose different phases for the long arm. The
probability for detection in the central time slot (corresponding to the two indistinguishable situations when the pump photon
has followed the short (long) arm of the interferometer and the two down
converted photons both the long (short) one \cite{Gisen} by a given
combination of detectors depends on the phases of the three interferometers involved in production and detection of the
photon pair \cite{Gisen} and on the polarisers' settings. Different choices
originate different detection bases.

Therefore, this scheme allows obtaining two independent (as the two
entanglement are independent) bits for each received photon, one related to
polarisation, the other to phase. When Alice and Bob have chosen the same
two bases (both for polarisation and phase) they have two correlated
outputs, which they can use for generating the key. The other choices can be
used for testing the channel (e.g. by means of Bell inequalities in the
Ekert's protocol).

As an example of security analysis of the channel let us consider
the simplest case where Eve decides to eavesdrop the photons
directed to Bob in one of the possible basis used by Alice and
Bob, both for the phase and the polarisation ones. The result of
this analysis \cite{epjd} is that if Eve intercepts a fraction
fraction $\eta$ of the transmitted photons, she  obtains an
information

\be
I_{AE} = 0.5 \eta
\ee

for the single entangled channel and

\be
I_{AE} = 0.25 \eta
\ee

for the double entangled one. In order to obtain the same information she is therefore bound to produce an error rate on the AB channel 3 times larger for the double entangled channel, making by far easier her identification in this case.
If she had chosen for
eavesdropping an intermediate basis (dubbed the Breidbart basis) for both
the phase and the polarisation ones, eavesdropping a fraction $\eta$ of the photons going to Bob, she would have obtained an
information $I_{AE} = 0.399 \eta $ for the single entangled channel and $%
I_{AE} = 0.189 \eta $ for the double entangled one. In order to obtain the
same information she should thus produce an error rate on the Alice-Bob
channel $19/6$ larger for the double entangled channel, which, as before,
results in a much larger chance of identifying the eavesdropping in the
double entangled case.

A similar analysis for other transmission schemes (as Eckert protocol) leads to analogous results \cite{epjd}.

\centerline{Conclusions}

In this proceeding I have presented a source of polarisation entangled photons realised at IENGF. This source is constituted of two $LiIO_3$ crystals whose type I PDC emission (of opposite polarization, having rotated the polarisation of the pump laser which pumps both of them) are superimposed by means of an optical condenser. Since the distance between the two crystals is smaller of the coherence length of the pump laser, the emitted state is an entangled one of the form:
 \be
\vert \Psi \rangle ={\vert H \rangle \vert H \rangle + f \vert V
\rangle \vert V \rangle \over \sqrt{1 + \vert f \vert ^2}}
\label{PsiH2}
\ee
where the degree of entanglement $f$ can be tuned by varying the pump intensity and phase between the two crystals. This source is rather brilliant: we measured a 10 kHz coincidence rate at 200mW pump power.

Because of these properties this source represents an interesting device for quantum communication and foundations of quantum mechanics experiments.

In this paper I have presented an experiment about Bell inequalities realised with it. Furthermore, I have hinted to an innovative double slit experiment that we have realised with a modification of this set-up.

Finally, I have discussed some future possible experiments in quantum communication to be realised by using this source.

\centerline{Acknowledgments} I would like to acknowledge all the
persons who collaborated to the work described in this paper: G.
Brida, E. Cagliero, G. Falzetta, M. Gramegna, E. Predazzi and C.
Novero, whom this paper is devoted {\it ad memoriam}.

I thank R. Steni for the realisation of the double slit.

I would like also to acknowledge support by  Italian minister of
University and Research MURST (contract 2001023718-002), by INTAS,
by Italian Space Agency (ASI, contract LONO-500172), by Italian
Institute of Nuclear Physics (INFN), by University of Turin  and
by Regione Piemonte.

\newpage
{\bf Figure Captions}

Fig. 1 Sketch of the source of polarisation entangled photons.
NLC1 and NLC2 are two $LiIO_3$ crystals cut
   at the phase-matching angle of $51^o$. L1 and L2 are two identical piano-convex lenses with a hole of 4 mm in the
   centre. C is a 5 x 5 x 5 mm quartz plate for birefringence compensation and $\lambda / 2$ is a first order half
   wave-length plate at 351 nm. U.V. identifies the pumping radiation at 351 nm.

Fig.2 Contour plot of the quantity $CH/N$ (see Eq. \ref{eq:CH}. N
is the total number of detections) in the plane with $f$ (non
maximally entanglement parameter, see the text for the definition)
as y-axis and $\eta$ (total detection efficiency) as x-axis. The
leftmost region (in black) corresponds to the region where no
detection loophole free test of Bell inequalities can be
performed. Ideal polarisers are considered.
 The contour lines are at 0, 0.01, 0.1, 0.15, 0.2.

 Fig.3 Set up for absolute calibration of photo-detectors by using PDC
\cite{Mig}. Two conjugated directions are sent to the detector
under calibration and to a second one (trigger), where a strong
spatial and band selection is performed. The number of single
counts on detector $i$ (i=1,2) is given by $N_i = \eta_i N$, where
$\eta_i$ is the quantum efficiency
 of the detector $i$ and $N$ the number of pairs. The coincidences number is $N_c = \eta_1 \eta_2 N$, thus the quantum
 efficiency $\eta_1$ of detector under calibration is given by $\eta_1 = N_c / N_2$. Corrections for accidental counts
  and coincidences and for electronic acquisition system must be added
  \cite{JMO}. For the sake of completeness, the alignment system
  is shown as well.

  Fig. 4  The set-up of the double slit experiment. A pump laser at 351 nm generates parametric down conversion of type I in a lithium-iodate crystal. Conjugated photons at 702
nm are sent to a double-slit by a system of two piano-convex
lenses in a way that each photon of the pair crosses a well
defined slit. A first photodetector is placed at  1.21 m  a second
one at 1.5 m from the slit. Both the single photon detectors (D)
are preceded by an interferential filter at 702 nm (IF) and a lens
(L) of 6 mm diameter and 25.4 mm focal length. Signals from
detectors are sent to a Time Amplitude Converter and then to the
acquisition system (multi- channel analyser and counters).

Fig. 5 Coincidences data in the region of interest compared with
quantum mechanics predictions.
 On the x-axis we report the position of the first detector respect to the median symmetry axis of the double slit.
 The second detector is kept fixed at -0.055 m (the region without data around this point is due to the superposition of the two detectors).
The x errors bars represent the width of the lens before the
detector. A correction for laser power  fluctuations is included.

Fig. 6 Plot of coincidences pattern (in arbitrary units) as a
function of the positions of the first
   photo-detector when the second one is kept fixed at $-1$ cm from the symmetry axis.

Fig. 7  Scheme for the generation of the double entangled photon
pairs. A Mach-Zender interferometer creates a state of the pump
photon which is given by the superposition of the states
corresponding to the photon following the long and the short path
respectively. The pump photon then generates or a horizontally
polarised pair in the first type I (NLC1) crystal either (after
having been rotated by a $\lambda / 2$ wave plate) a vertically
polarised one in the second type I (NLC2) crystal. The parametric
down conversions of the two crystals are then superimposed using
an optical condenser with a hole drilled in the centre for leaving
pass the pump undisturbed. The optical path of idler, signal and
pump are arranged by means of compensator elements (C) for not
introducing any delay among these. The superposition of the
probability of generating a pair in the first or in the second
crystal originates the polarisation entanglement.

Fig. 8 The scheme for the reception apparatus of Alice and Bob. A
prism, properly rotated, allows a polarisation selection. On each
arm exiting the prism a Mach-Zender interferometer is inserted
with a phase shift on the long arm which is suitably arranged by
the observer. Photo-detectors are denoted by an ellipse.

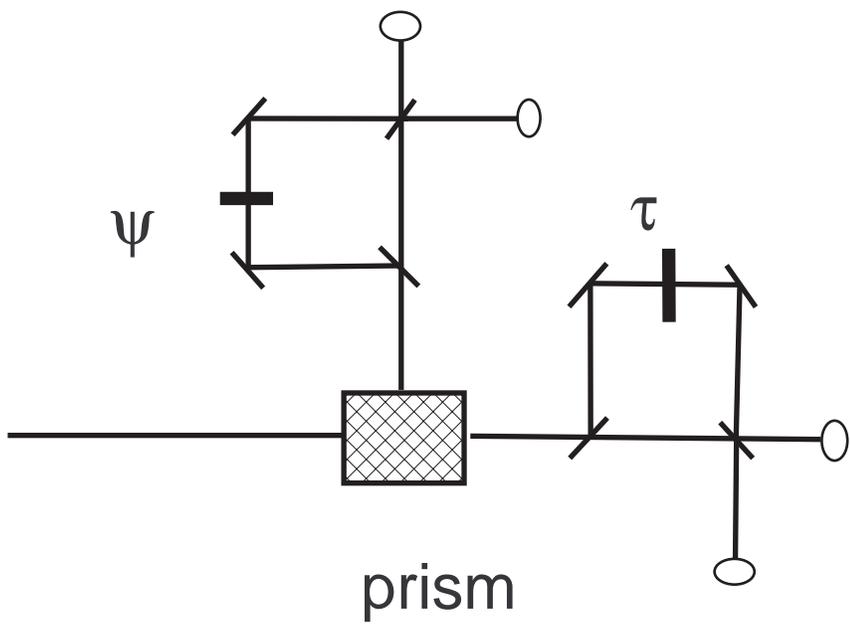
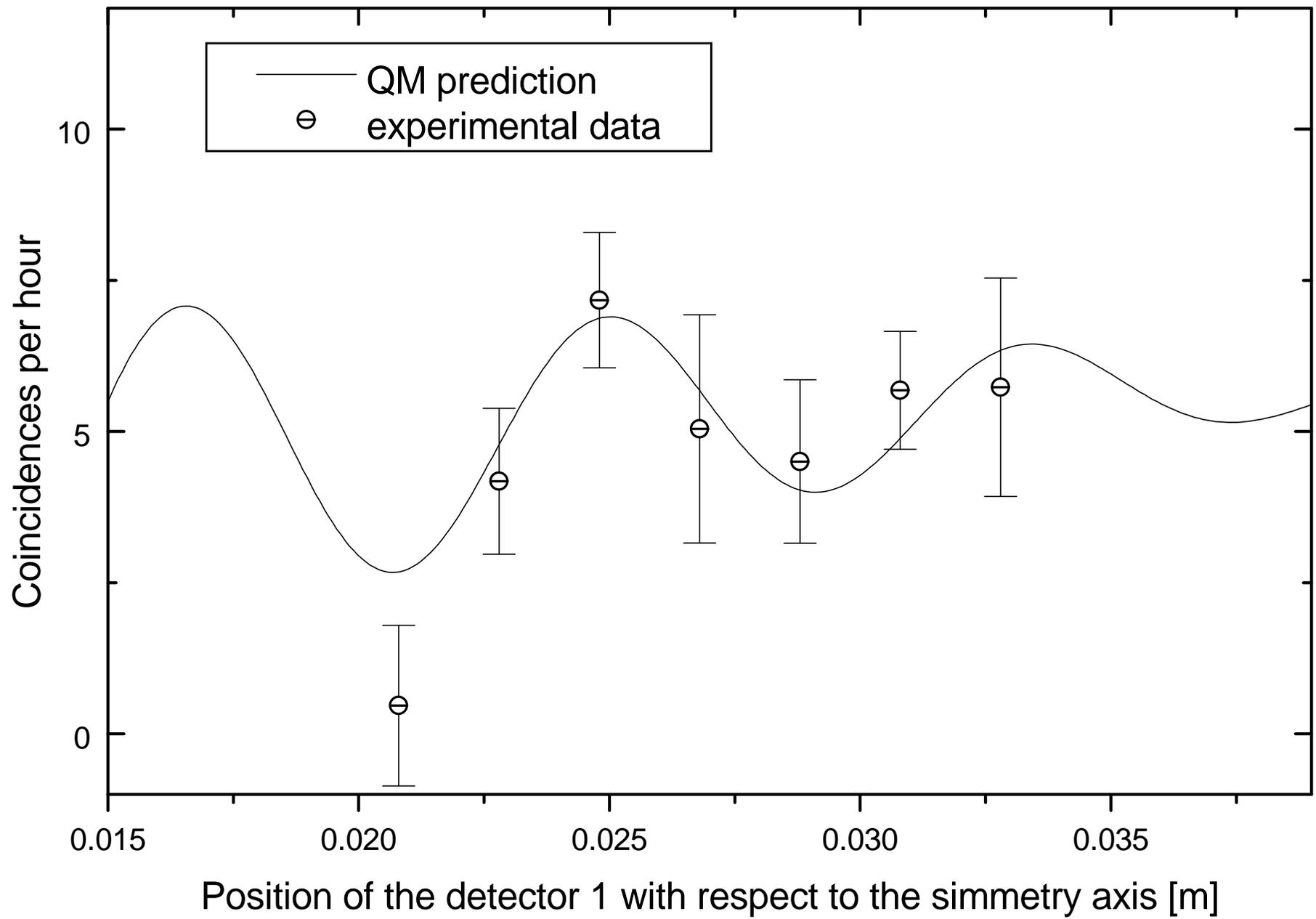
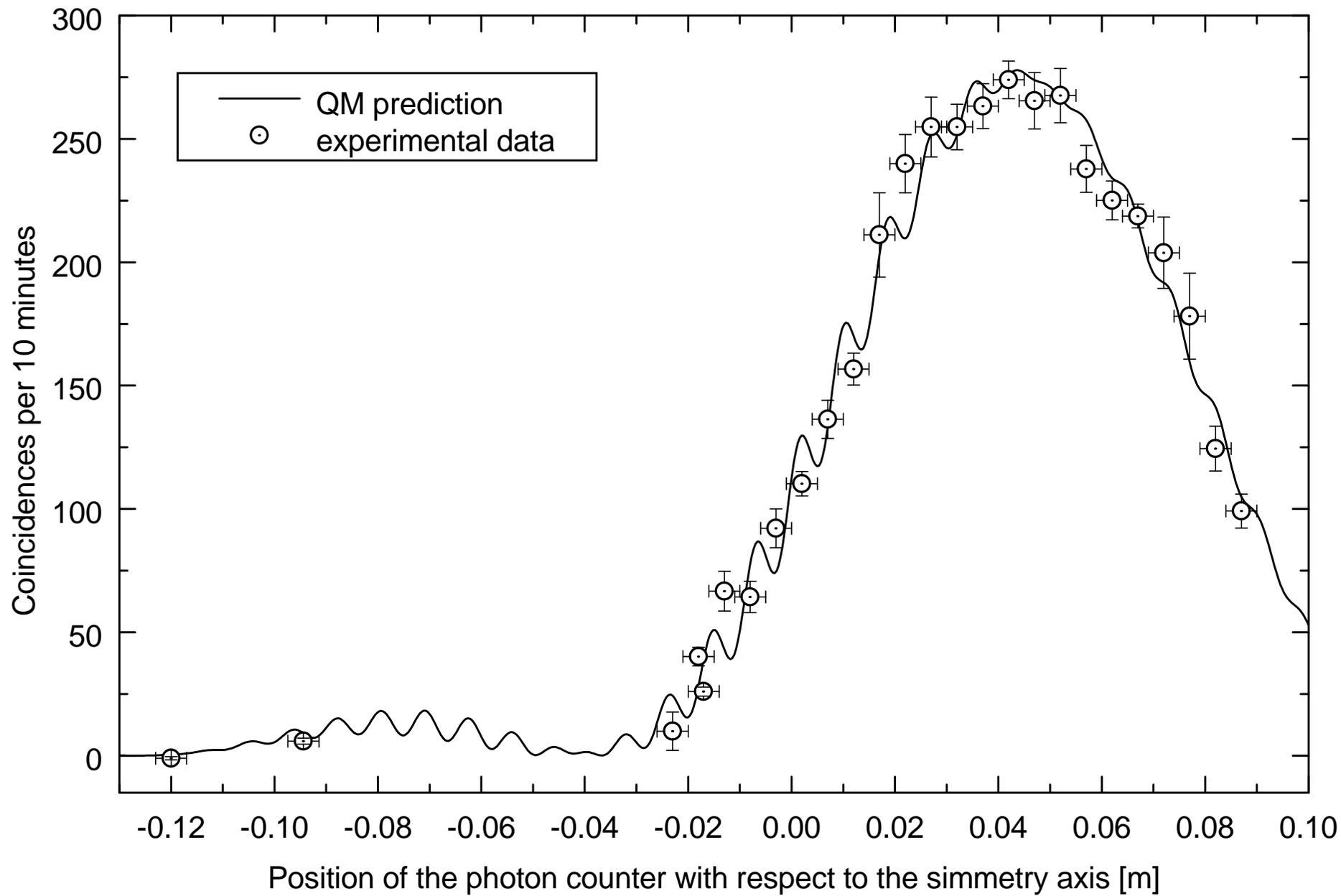
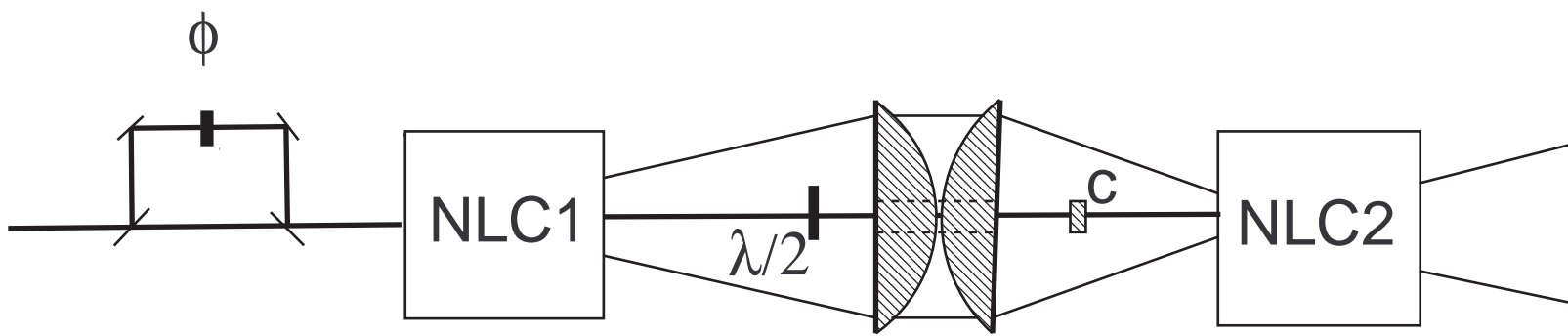
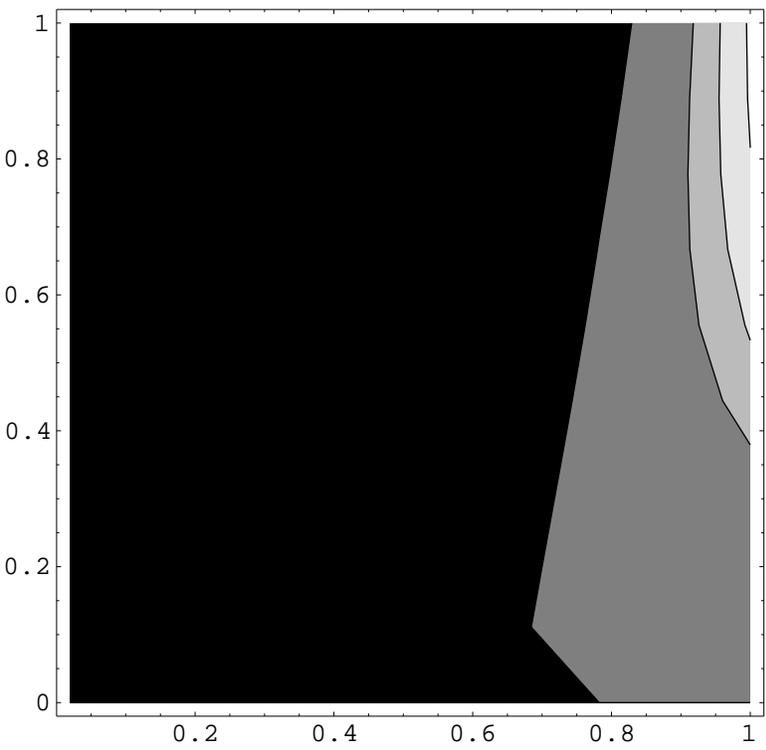
\end{document}